\documentclass[aps, prb, preprint, bibnotes, showpacs, preprintnumbers, amsmath, amssymb, superscriptaddress]{revtex4}

\usepackage{graphicx}% Include figure files
\usepackage{dcolumn}% Align table columns on decimal point
\usepackage{bm}% bold math
\usepackage{amssymb}

\begin{document}

\title{Large room-temperature magnetoresistance in lateral organic spin valves fabricated by in-situ shadow evaporation}

\author{M.~Gr\"{u}newald}
\affiliation{Physikalisches Institut (EP3), Universit\"{a}t W\"{u}rzburg, Am Hubland, D-97074
W\"{u}rzburg, Germany} \affiliation{Institute of Physics, Martin-Luther-Universit\"{a}t
Halle-Wittenberg, Von-Danckelmann-Platz 3, D-06120 Halle, Germany}

\author{J.~Kleinlein}
\affiliation{Institute of Physics, Martin-Luther-Universit\"{a}t
Halle-Wittenberg, Von-Danckelmann-Platz 3, D-06120 Halle, Germany}

\author{F.~Syrowatka}
\affiliation{IZM, Martin-Luther-Universit\"{a}t Halle-Wittenberg, D-06099, Germany}

\author{F. W\"{u}rthner}
\affiliation{Wilhelm Conrad R\"{o}ntgen Research Center for Complex Material Systems, Universit\"{a}t W\"{u}rzburg, Am Hubland, 97074 W\"{u}rzburg, Germany} \affiliation{Institut f\"{u}r Organische Chemie,
Universit\"{a}t W\"{u}rzburg, Am Hubland, D-97074 W\"{u}rzburg, Germany}

\author{L.W. Molenkamp}
\affiliation{Physikalisches Institut (EP3), Universit\"{a}t W\"{u}rzburg, Am Hubland, D-97074
W\"{u}rzburg, Germany}

\author{G. Schmidt}
\email[Correspondence to G. Schmidt: ]{georg.schmidt@physik.uni-halle.de}
\affiliation{Physikalisches Institut (EP3), Universit\"{a}t W\"{u}rzburg, Am Hubland, D-97074
W\"{u}rzburg, Germany} \affiliation{Institute of Physics, Martin-Luther-Universit\"{a}t
Halle-Wittenberg, Von-Danckelmann-Platz 3, D-06120 Halle, Germany}\affiliation{IZM, Martin-Luther-Universit\"{a}t Halle-Wittenberg, D-06099, Germany}


\begin{abstract}
We report the successful fabrication of lateral organic spin valves with a channel length in the sub $100\,nm$ regime. The fabication process is based on in-situ shadow evaporation under UHV conditions and therefore yields clean and oxygen-free interfaces between the ferromagnetic metallic electrodes and the organic semiconductor. The spin valve devices consist of Nickel and Cobalt-iron electrodes and the high mobility \emph{n}-type organic semiconductor $N,N'$-bis(heptafluorobutyl)-$3,4:9,10$-perylene diimide. Our studies comprise fundamental investigations of the process' and materials' suitability for the fabrication of lateral spin valve devices as well as magnetotransport measurements at room temperature. The best devices exhibit a magnetoresistance of up to $50\,\%$, the largest value for room temperature reported so far.

\end{abstract}

\maketitle

\section{Introduction}
Lateral organic spin valves (OSVs)\cite{Dediu2002} hold the promise of integrating non-volatile switching into organic field effect transistors (OFETs). In contrast to vertical spin valves, however, they are much more difficult to fabricate. While in vertical spin valves the thickness of the organic layer determines the spacing between the magnetic electrodes\cite{Xiong2004} which can be as small as a few nanometers, lateral spin valves need highly sophisticated patterning to achieve a similarly small gap between the two contacts. This patterning usually involves electron-beam lithography based processing \cite{Michelfeit2008,Golmar2012} resulting in massive contamination of the contact surfaces and thus impeding spin valve operation.
We have developed a novel fabrication process for lateral spin valves with channel lengths of less than $100\,nm$ and contamination free interfaces.

Two prerequisites are mandatory for an organic spin valve: firstly, two ferromagnetic electrodes must be prepared that have different coercive fields in order to allow for parallel or antiparallel magnetization of the electrodes. Secondly, the spacing between the two electrodes which is filled with the organic semiconductor (OSC) must be smaller than the spin diffusion length. Any reports so far show that the gap should not exceed $100\,nm$ in order to allow for sizeable spin valve effects\cite{Xiong2004,Santos2007,Nguyen2010}.
In vertical spin valves these preconditions can easily be met by depositing a stack of two different ferromagnetic materials with an OSC of suitable thickness in between. Clean interfaces can be obtained by depositing the whole layer stack in a UHV chamber. For lateral devices the situation is more complicated. A lateral gap of less than $100\,nm$ is usually achieved by electron beam patterning of a ferromagnetic layer either using etching or evaporation and lift-off\cite{Michelfeit2008,Golmar2012}. The coercive fields of the contacts can be tuned in the same process step by making striped contacts of different widths in the sub-micron range\cite{Michelfeit2008,Jedema2001}. This patterning process, however, results in a massive contamination and typically oxidation of the surface of the ferromagnetic contacts before the OSC is deposited. Resistless patterning using a solid shadow mask without the need for a lift-off can circumvent this problem, however, the lateral resolution of mechanically fabricated masks does not reach down to the necessary $100\,nm$ range.

\section{Material and methods}
\begin{figure}

\includegraphics[]{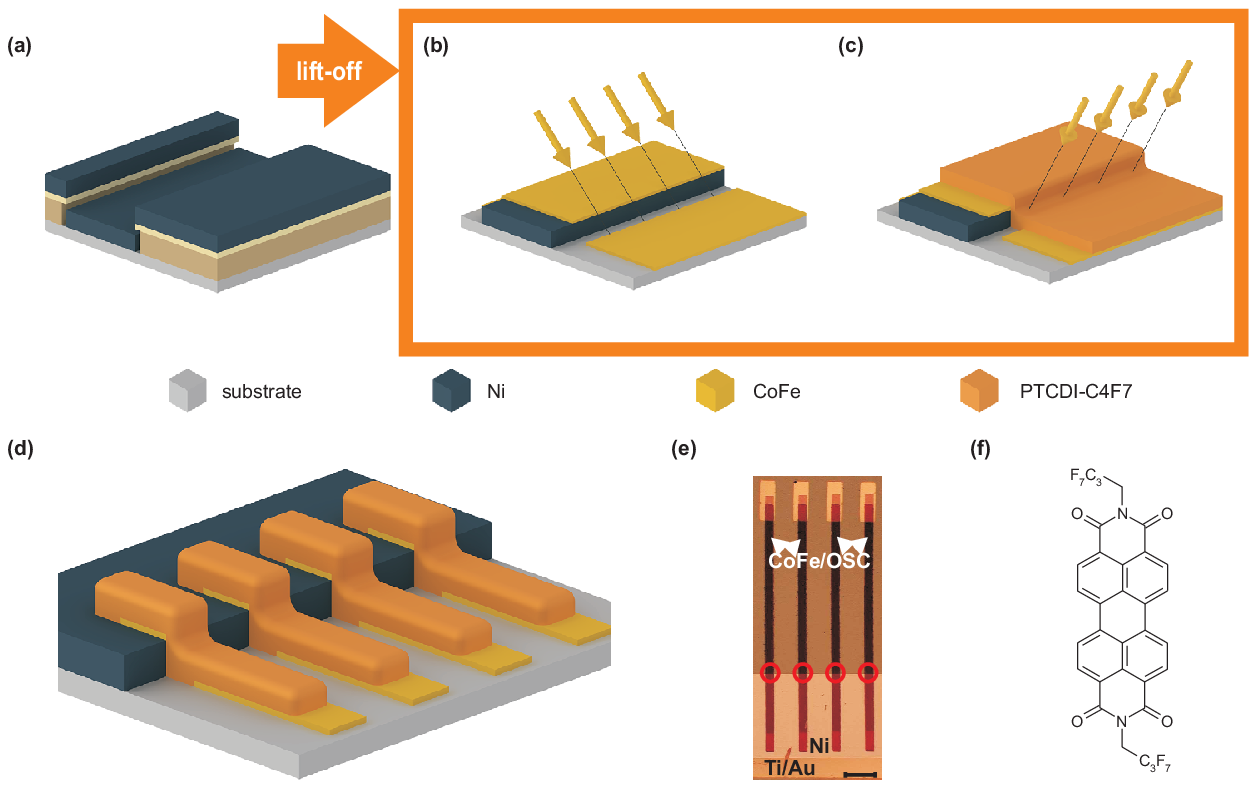}

\caption{\textbf{Novel fabrication process for lateral organic spin valve structures.} (a) Patterning of the thick Ni layer by optical lithography and lift-off. (b, c) The deposition of the second electrode  and the evaporation of the conducting OSC layer are done in-situ yielding clean interfaces between electrodes and OSC. Uniform device dimension are obtained by covering the sample with a shadow mask with rectangular windows. (d) Schematic drawing of single devices. (e) Microscopic photograph of a sample with four devices. Titanium/Gold contact pads for ultrasonic bonding are patterned prior to the described fabrication process. The gaps are marked with red circles. (f) Molecular structure of the OSC PTCDI-C4F7. \label{FigFab}}
\end{figure}

In our process we are using a combination of optical lithography, lift-off, and shadow evaporation under different angles. This process yields uncontaminated interfaces, a small gap for the OSC and contacts with a sizeable difference in coercive field (flow chart shown Fig. \ref{FigFab}a-c).

In the first step a rectangular stripe of Ni (0.75$\,mm$ x 4.0$\,mm$) is patterned by optical lithography, evaporation, and lift-off on a silicon wafer covered by $1\,\mu m$ of silicon oxide (Fig. \ref{FigFab}a). Here, vertical sidewalls of the Ni are crucial for the subsequent steps. Without any further treatment, the sample is then transferred into a UHV chamber (base pressure $1\cdot10^{-9}\,mbar$) in which first a thin layer of CoFe is evaporated with the sample tilted by $\Theta_{evaporation}=45\,^\circ$ (Fig. \ref{FigFab}b). The shadow of the Ni contact results in a gap of less than $100\,nm$ which now separates two contact areas: one bilayer contact (Ni and CoFe) and a single layer contact (CoFe only). The nominal length of the separating gap $l_{channel}$ is determined by the thickness of the Ni layer $t_{Ni}$ as:

\begin{equation}
l_{channel}=t_{Ni}\cdot tan(\Theta_{evaporation})
\end{equation}

Without breaking the UHV-conditions the gap is then filled by evaporation of the \emph{n}-type OSC PTCDI-C4F7 ($N,N'$-bis(heptafluorobutyl)-$3,4:9,10$-perylene diimide, molecular structure shown in Fig. \ref{FigFab}f) under a suitable angle (Fig. \ref{FigFab}c). PTCDI-C4F7 is used due to its excellent properties with respect to stability in ambient conditions and electron mobility \cite{Oh2007,Gruenewald2011}. According to \citet{Oh2007} the stability of PTCDI-C4F7 layers originates from their high packing density prohibiting oxygen and moisture diffusion. Hence the OSC layer in our devices also protects the underlying ferromagnetic electrodes from contamination after removal from the UHV-chamber.\\
During the two in-situ evaporation steps a large area shadow mask with rectangular windows is used to achieve well separated and uniform devices (Fig. \ref{FigFab}d and e) without the need for any further lithography. Typical dimensions of the devices are: channel width $w_{channel}\approx 150\,\mu m$, thickness of the CoFe contact $t_{CoFe}=10\,nm$, $t_{Ni}=80\,nm$ and a resulting $l_{channel}\approx 80\,nm$. \\

\section{Results and discussion}
\begin{figure}

\includegraphics[]{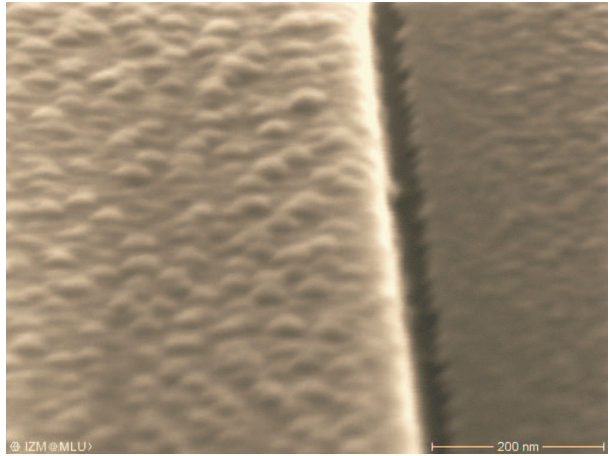}

\caption{\textbf{Scanning Electron Microscopy image of a sub $\textbf{100\,nm}$ channel fabricated by shadow evaporation.} A $80\,nm$ thick Ti/Au layer (left hand side, light grey) is used for shadow evaporation with the sample tilted by $\Theta_{evaporation}=45\,^\circ$. The second contact is a $10\,nm$ thick Ti layer (right hand side, dark grey). The length of the channel, which is varying due to small protrusions on the side of the thin layer, is approximately only half of the theoretical value of $80\,nm$.\label{FigESEM}}
\end{figure}

Fig. \ref{FigESEM} shows a scanning electron microscope (SEM) image of the gap of a sample device after the first tilted metal evaporation. Au and Ti were used instead of Ni and CoFe in order to enhance the contrast. Obviously the nominal channel length of $80\,nm$ is not achieved, most likely due to diffusion and slightly tilted sidewalls.

Furthermore the channel length is not constant but is varying due to imperfections at the edge of the contact defined by optical lithography. The shape of the Au edge is transferred also to the shape of the Ti layer leading to small protrusions of different sizes.\\

\begin{figure}

\includegraphics[]{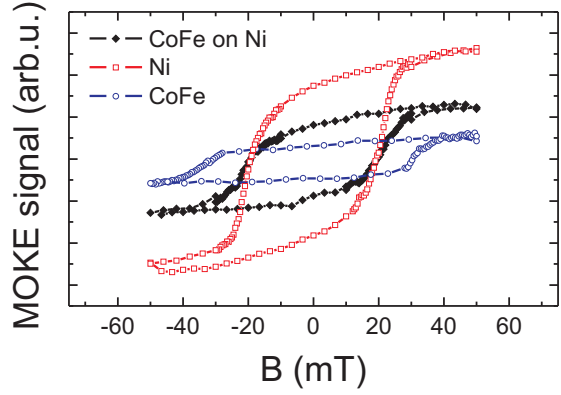}

\caption{\textbf{MOKE measurements at room temperature from extensive magnetic layers.} The layer compositions for the black (diamonds) and blue (open circles) traces are similar to those of the devices' contacts: $10\,nm$ CoFe on top of $80\,nm$ Ni (black, diamonds) and $10\,nm$ CoFe (blue, open circles). The red curve (open squares) shows the hysteresis of a Ni layer with $80\,nm$ thickness indicating that for the bilayer contact the magnetization of the CoFe layer is pinned to that of the Ni layer. \label{FigMoke}}
\end{figure}

The magnetic properties of the devices are studied by Magneto optical Kerr rotation (MOKE) at room temperature (Fig. \ref{FigMoke}). Measurements are performed on reference samples with layers of the following compositions:  a single $80\,nm$ thick Ni layer (red curve, line with open squares), a bilayer ($10\,nm$ CoFe on top of $80\,nm$ Ni, black curve, line with diamonds), and a $10\,nm$ CoFe layer (blue curve, line with open circles). The last two layers correspond to our two device contacts. Due to the large penetration depth of the probe beam the MOKE signal is a superposition of the signals collected from CoFe and Ni, respectively. The measurements show that for the bilayer the two metals are ferromagnetically coupled and the (thicker) Ni is dominating the switching ($H_{c, bilayer}\,\approx\,20\,mT$), while a single CoFe layer has a considerably higher coercive field ($H_{c, CoFe}\,\approx\,30\,mT$).
Therefore also the coercive fields of our devices' contacts are different, one decisive prerequisite for true spin valve operation.

\begin{figure}

\includegraphics[]{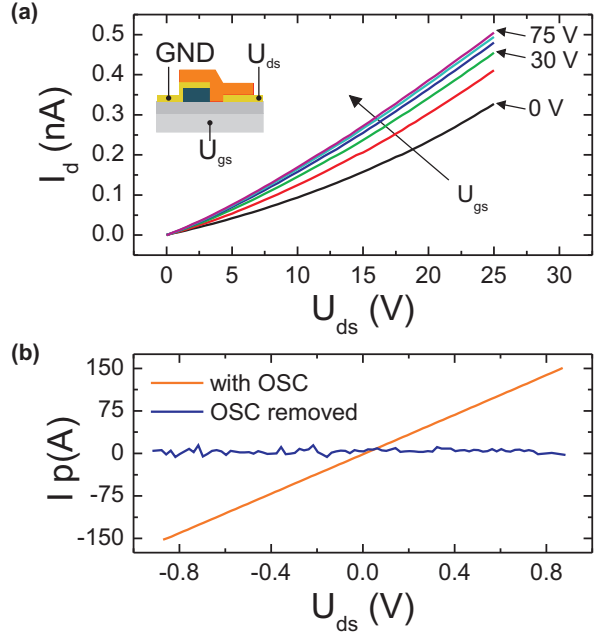}

\caption{\textbf{Results from transport measurements showing the electrodes' separation.} (a) Typical output characteristics of a spin valve device fabricated by shadow evaporation: the substrate is used as gate electrode (applied constant gate voltage $U_{gs}$), the drain source voltage $U_{ds}$ is applied between the electrodes as shown in the insert. Short channel effects are visible in these measurements, such as the absence of saturation of the drain current $I_{d}$ at high $U_{ds}$. (b) Additional proof of the electrodes' separation: having recorded the I/V characteristic of another spin valve structure (light grey/orange line) the OSC layer was removed and the measurement repeated (black/blue line): the charge transfer between the electrodes can be clearly attributed to the OSC layer and short circuits or weak metallic links therefore be excluded. All measurements are performed at room temperature.\label{FigGate}}
\end{figure}

Electrical separation of the two contacts is checked by using them as drain and source of an OFET and the substrate as the gate electrode (insert of Fig. \ref{FigGate}a). Output characteristics $I_d(U_{ds})$ at constant $U_{gs}$ ($I_d$ drain current, $U_{ds}$ drain source voltage, $U_{gs}$ gate source voltage) in Fig. \ref{FigGate}a show a clear dependence of $I_d$ on $U_{gs}$. However, neither saturation of $I_d$ nor the typical quadratic dependence of $I_{d}$ on $U_{gs}$ can be observed at high $U_{ds}$. With a channel length much smaller than the oxide thickness the gradual channel approximation is no longer valid \cite{Tsividis1999,Sze2006,Pierret1996,Haddock2006} and the OFETs are in the short channel regime. In a second control experiment (Fig. \ref{FigGate}b) the OSC is removed after the I/V-characterization, leading to a channel current below the measurement limit, a clear indication that the only current path is through the OSC.

As can be seen from the I/V and output characteristics in Fig. \ref{FigGate}b the devices exhibit large resistances in the range of a few $G\Omega\,s$ to hundreds of $G\Omega\,s$ at room temperature. The observed drain currents in the range of $I_d\approx\,1-10\,pA$ at low $U_{ds}$ are in good agreement with unpublished studies of PTCDI-C4F7-based OFET structures with ferromagnetic metal electrodes which were patterned by optical lithography ($l_{channel}=1.5-10\,\mu m$). Because with decreasing temperature the current drops below the measurement limit, all magnetotransport studies are carried out at room temperature.

\begin{figure}

\includegraphics[]{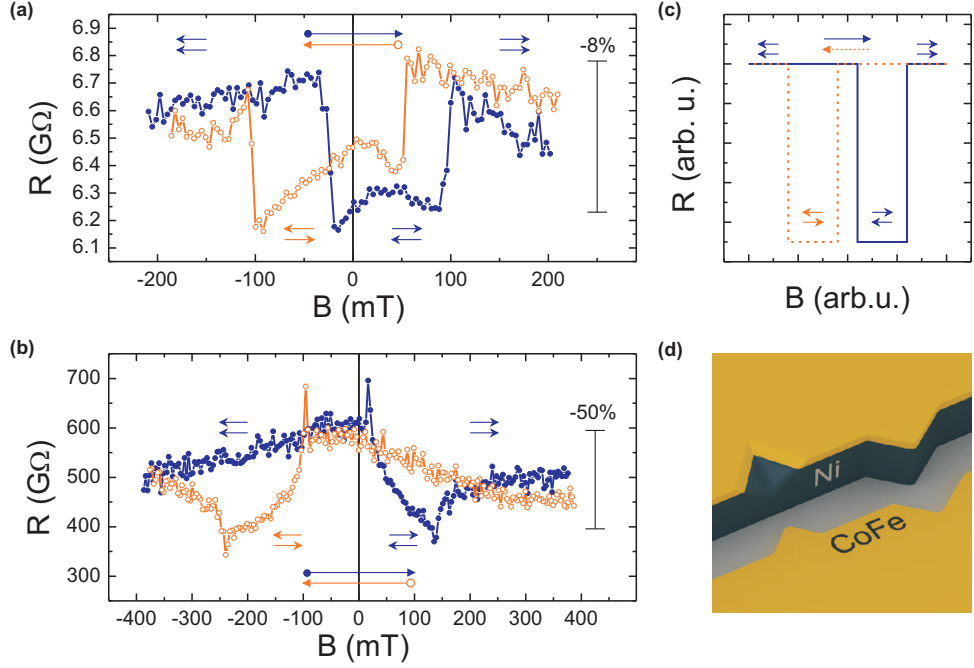}

\caption{\textbf{Typical magnetoresistance signals of two different lateral spin valve structures recorded at room temperature.} (a,b) The big arrows in the bottom/top of the panels indicate the sweep direction for the blue and orange curve (curve with closed/open circles respectively). The small arrows near by the curves correspond to the relative magnetization of the two electrodes assuming spin valve behavior. (a) Spin valve like signal with a ratio of $8\,\%$ at a low device resistance, recorded at $U_{ds}=50\,mV$. (b) Spin valve like signal with a ratio of $50\,\%$ ($90\,\%$ when considering the sharp peaks) at a high device resistance, recorded at $U_{ds}=1\,V$. (c) Schematic of the normal negative spin valve effect of a device with macroscopic ferromagnetic electrodes. The magnetization of the electrodes is assumed to be completely reversed at a certain field. (d) Schematic drawing of protrusions on the ferromagnetic electrodes and possible configurations for a lateral tunneling junction. \label{FigMR}}
\end{figure}

Fig. \ref{FigMR}a and b show the magnetoresistance (MR) traces of two different spin valve structures. For both devices we observe a hysteretic spin valve behavior with a negative magnetoresistance which is often observed in OSV devices. Shape and magnitude of the two curves differ, however, considerably and cannot be explained by the simple reversal of two ferromagnets.
Device a has a baseline resistance of $6.6\,G\Omega$ and shows two distinct switching events for each sweep direction. However, when increasing the magnetic field from negative saturation towards positive values the first magnetization reversal is observed before zero B-field is reached which is in contrast to normal spin valve behavior (Fig. \ref{FigMR}c).  In this device the magnetoresistance $\Delta R/R_{anti-parallel}$ is $8\,\%$ (absolute value).
For device b the baseline resistance is $\approx 500\,G\Omega$. The magnetoresistance trace indicates a magnetization rotation rather than a switching. The absolute magnitude of the magnetoresistance is increased to $50\,\%$. In device b the MR trace is also hysteretic, however, with a lateral shift reminiscent of exchange bias. With the Ni electrode being exposed to air during the processing, it is indeed likely that NiO, which is an antiferromagnet with a N\'{e}el temperature well above room temperature, leads to exchange bias in the electrode\cite{CRC1996}.

It should be noted that the characteristic magnetic fields in both MR traces are higher than would be expected from the MOKE characterization of the contact layers.

While exchange bias only is not sufficient to explain the shape of the MR traces it can be understood if the possible transport mechanism is reconsidered based on findings of \citet{Barraud2010}.

\citet{Barraud2010} have demonstrated the fabrication of a nanometer-sized pin-hole like artificial TMR contact through an OSC using conducting atomic force microscopy. The baseline resistance that they observe is in a similar regime as the one seen in our experiments. Also the magnetoresistance traces are more complex than typical for OSV structures.

We suspect that in our devices the conductance channel is a very small lateral tunneling contact created by a small protrusion on one of the two ferromagnetic contacts (Fig. \ref{FigMR}d). The resistances of our contacts are in the regime observed by Barraud et al., while the variations from device to device are large corresponding to statistical size variations of accidental protrusions. Large area current injection over the whole contact edge, however, would result in a resistance proportional to the gap size, and variations from device to device should be small.

This concept is also helpful to understand the shape of the magnetoresistance traces which is now largely dominated by the local domain structure of the small tip. In addition, the tip is in direct vicinity of the opposing ($mm$-sized) contact which is less than $100\, nm$ away. Because the magnetic field is applied along the stripes (crossing the gap in the direction from one contact to the other) the stray field is extremely large. Hence we expect to observe a strong dipolar coupling between the tip and the opposing contact resulting in switching fields different from those observed by MOKE.

\section{Conclusion}
We have demonstrated that a simple shadow evaporation process can be used for the fabrication of lateral OSVs with clean interfaces and different switching fields of the ferromagnetic contacts. The devices show a large magnetoresistance in the range of a few $\%$ up to $50\,\%$ at room temperature. The observed effect can be explained as a spin valve signal in a lateral tunnel junction where one of the electrodes is a small, randomly shaped protrusion on a ferromagnetic contact.

\section{Acknowledgements}
We thank the EU for funding the research in the projects OFSPIN (NMP-CT-2006-033370) and HINTS (NMP3-SL-2011-263104).

\clearpage

\clearpage

\clearpage

\clearpage

\clearpage

\end{document}